\documentclass[conf]{new-aiaa}
\usepackage[utf8]{inputenc}
\usepackage{graphicx}
\usepackage{subcaption}
\usepackage{amsmath}
\usepackage[version=4]{mhchem}
\usepackage{siunitx}
\usepackage{nicefrac}
\usepackage{longtable,tabularx}
\hypersetup{hidelinks}
\title{Generating Physically Plausible Parachute Dynamics with Deep Generative Modeling}
\author{Yulong Yang
        \footnote{PhD Candidate, Mechanical and Aerospace Engineering, Engineering Quadrangle, Princeton, NJ 08544.}}
\affil{Princeton University, Princeton, New Jersey}
\author{Clara O'Farrell
       \footnote{Group Supervisor, Dynamics, Modeling, Flight Mechanics \& UQ, 4800 Oak Grove Dr, M/S 198-326, Pasadena, CA 91011, AIAA Member.}}
\affil{Jet Propulsion Laboratory, California Institute of Technology, Pasadena, California}
\author{Christine Allen-Blanchette
        \footnote{Assistant Professor, Mechanical and Aerospace Engineering, Engineering Quadrangle, Princeton, NJ 08544.}
        \footnote{Assistant Professor, Center for Statistics and Machine Learning, Engineering Quadrangle, Princeton, NJ 08544.}}
\affil{Princeton University, Princeton, New Jersey}
\begin{document}
\maketitle
\begin{abstract}
    Accurately modeling the dynamics of planetary parachute and entry vehicle systems is critical for Entry, Descent, and Landing events such as vehicle separation and sensor activation. These dynamics are difficult to capture with traditional system-identification methods as parachute motion is highly nonlinear, the governing equations are not fully known, and relevant test data are scarce and expensive to acquire. In this work, we sidestep these challenges by leveraging a physics-aware generative modeling approach that learns parachute dynamics directly from data. The proposed method, Symplectic Parachute Generative Adversarial Network (SPar-GAN), adapts a Hamiltonian generative architecture to the parachute setting by conditioning on canopy design and freestream velocity, while enforcing conservation of energy through symplectic integration. We apply SPar-GAN to subscale parachute tests conducted at the National Full-Scale Aerodynamics Complex and show that it reproduces qualitatively accurate pitch–yaw dynamics of different parachute configurations while recovering a compact two-degree-of-freedom phase-space consistent with canopy axisymmetry. These results suggest that physics-constrained generative models can characterize parachute dynamics across operating conditions and may help reduce the volume of physical testing required to assess performance.
\end{abstract}
\section{Introduction}
%
\lettrine{D}{eveloping} models that accurately capture the dynamics of planetary parachute and entry vehicle (EV) systems has long proven challenging owing to the scarcity of data in relevant environments, the nonlinear nature of the system, and uncertainties about the form of the underlying equations of motion. Existing models have had great success in capturing performance metrics such as inflation loads and steady state drag coefficient. However, past attempts have been less successful in capturing the long-term dynamics of the parachute-EV system~\citep{cruz2014reconstruction,o2024reconstructed}. These dynamics are of interest because events critical to the success of the Entry, Descent, and Landing (EDL) sequence, such as separation events or sensor activation~\citep{johnson2023implementation}, often take place during parachute terminal descent. To date, our approach has been to design EDL systems conservatively, at the expense of mass and fuel margins. As we advance towards increasing our landed mass for crewed missions to Mars and consider novel EDL architectures~\citep{shapiro2024dragonfly}, understanding these dynamics in order to reduce our design margins becomes increasingly important.

Equations of motion (EOM) for the dynamics of axisymmetric parachutes were developed in the 1960’s by extending the EOM for aircraft linearized about steady level flight~\citep{white1968theory,wolf1971strability}. However, extracting the required aerodynamic coefficients from experimental data using traditional systems identification techniques has proven challenging. Recent attempts using data captured during subscale wind tunnel campaigns~\citep{schoenenberger2005parachute, gonyea2013aerodynamic} have had limited success, as they require significant domain knowledge to accurately model system behavior.  Depending on the method, this knowledge may take the form of an analytical model of the dynamics~\citep{magal2018parameter,galioto2020bayesian}, a general functional form within which the dynamics are assumed to lie~\citep{epperlein2015thermoacoustics,paredes2024output}, or a curated library of candidate nonlinear terms from which the governing equations are selected~\citep{brunton2016discovering}.

Deep generative models offer a compelling alternative, having shown success in modeling systems with ambiguous dynamics based solely on example trajectories~\citep{bond2021deep}. A limitation of these models, however, is that they approximate the data distribution with a black-box neural network and offer no guarantee that generated trajectories respect physical conservation laws. Several recent approaches address this by incorporating physically motivated constraints such as Hamilton's equations directly into the generative model, so that generated trajectories are constrained to conserve relevant quantities~\citep{toth2019hamiltonian,allen2024hamiltonian,liu2025physically}. Building on recent work by~\citet{liu2025physically}, we develop Symplectic Parachute Generative Adversarial Network (SPar-GAN), which adapts the Hamiltonian generative framework to the parachute setting. Parachute dynamics present several challenges that distinguish them from the simulated canonical mechanical systems studied in previous work, including learning from noisy real-world measurements, operating on partial observations, and recovering a phase-space representation without a known reference dimensionality. Despite these challenges, we show that SPar-GAN reproduces accurate pitch-yaw dynamics of multiple distinct canopy designs, recovers a compact phase-space structure consistent with canopy axisymmetry, and generalizes across wind tunnel freestream velocities.

This paper is organized as follows. In Section~\ref{sec:background}, we review the relevant background on Hamiltonian mechanics, Hamiltonian Neural Networks, and Generative Adversarial Networks. Section~\ref{sec:SParGAN} describes the SPar-GAN architecture, including the configuration space map, phase-space motion model, and trajectory generation procedure. Section~\ref{sec:dataset} introduces the wind tunnel dataset used in our experiments. Section~\ref{sec:results} presents results on three canopy types -- ringsail, disksail, and modified DGB -- as well as conditional interpolation, extrapolation, and transfer experiments. Finally, Section~\ref{sec:conclusion} concludes with a summary of SPar-GAN and directions for future work.
%
\section{Background}\label{sec:background}
%
SPar-GAN leverages ideas from Hamiltonian mechanics, neural network-based dynamical systems, and adversarial generative modeling.
In this section, we briefly review each of these building blocks; specifically, Hamiltonian mechanics and the properties we exploit for phase-space recovery (Section~\ref{subsec:hamilton_mechanics}), Hamiltonian Neural Networks and symplectic integration for energy-consistent trajectory rollouts (Section~\ref{subsec:hamilton_neural_network}), and Generative Adversarial Networks for learning trajectory distributions conditioned on system parameters (Section~\ref{subsec:gan}).
%
%
\subsection{Hamiltonian Mechanics}\label{subsec:hamilton_mechanics}
Hamiltonian mechanics is a reformulation of classical mechanics that describes the evolution of a system in terms of energy rather than forces.
The Hamiltonian of a system $\mathcal{H}$ maps the generalized position and momentum coordinates $\left(\mathbf{q}, \mathbf{p}\right)$ to the total energy, and the dynamics of the system are governed by Hamilton's equations
\begin{equation}
    \dot{\mathbf{q}} = \frac{\partial\mathcal{H}}{\partial \mathbf{p}}, \quad\quad \dot{\mathbf{p}} = -\frac{\partial\mathcal{H}}{\partial \mathbf{q}}.\label{eq:hamilton_equation}
\end{equation}
A useful property of this formulation is that generalized coordinates which do not appear in the Hamiltonian do not contribute to the total energy and their conjugate momenta are conserved. 
Such coordinates are called cyclic or ignorable, and satisfy
\begin{equation}
    \dot{p}_{i}=-\frac{\partial\mathcal{H}}{\partial q_{i}}=0.
\end{equation}
We exploit this property in Section~\ref{subsec:configuration_space_map} to encourage SPar-GAN to recover both a set of canonical coordinates and the degrees of freedom of the parachute system.
%
%
\subsection{Hamiltonian Neural Network}\label{subsec:hamilton_neural_network}
Although Hamiltonian mechanics provides a general framework for describing conservative dynamical systems, the system Hamiltonian is rarely available in closed form for complex engineering systems such as parachutes, where aerodynamic interactions and structural deformation make analytical derivation impractical.
Hamiltonian Neural Networks (HNN)~\citep{greydanus2019hamiltonian} address this problem by replacing the analytical Hamiltonian with a learned one, parameterizing $\mathcal{H}\left(\mathbf{q}, \mathbf{p}\right)$ as a multilayer perceptron (MLP)~\citep{rumelhart1986learning} and fitting its parameters to trajectory data.

HNNs are trained by aligning the gradients of the learned Hamiltonian with the observed time derivatives of position and momentum.
Given trajectory samples $\left(\mathbf{q},\mathbf{p},\dot{\mathbf{q}},\dot{\mathbf{p}}\right)$, the HNN loss
\begin{equation}
    \mathcal{L}_{\mathrm{HNN}} = \left\lVert \frac{\partial\mathcal{H}}{\partial\mathbf{p}} - \dot{\mathbf{q}}\right\rVert_{2} + \left\lVert \frac{\partial\mathcal{H}}{\partial\mathbf{q}} + \dot{\mathbf{p}}\right\rVert_{2},
\end{equation}
where the partial derivatives are obtained by automatic differentiation. The loss function itself penalizes any trajectory that departs from ground truth observations.
Once trained, the learned Hamiltonian produces trajectories from any initial condition by integrating Hamilton's equations forward in time. 

Although HNNs respect energy conservation in principle, numerical integration of the learned dynamics introduces error that accumulates over long rollouts.
Standard explicit schemes such as forward Euler do not preserve the symplectic structure of Hamilton's equations and typically produce trajectories that drift in total energy.
Symplectic integrators avoid this by preserving a discrete analog of the symplectic structure of Hamilton's equations, which bounds the energy error over arbitrarily long rollouts.
One choice of symplectic integrator is leapfrog integration~\citep{chen2019symplectic}, a second-order scheme that we adopt in this work.
Leapfrog requires that the Hamiltonian is separable into kinetic and potential components,
\begin{equation}
    \mathcal{H}\left(\mathbf{q}, \mathbf{p}\right) = T\left(\mathbf{p}\right) + V\left(\mathbf{q}\right),
\end{equation}
in which case one step of integration consists of a half-step momentum update, a full-step position update, and a second half-step momentum update,
\begin{align}
    \mathbf{p}_{n+\nicefrac{1}{2}} &= \mathbf{p}_{n} - \nicefrac{1}{2}\Delta t T'\left(\mathbf{q}_{n}\right),\label{eq:leapfrog_1}\\
    \mathbf{q}_{n+1} &= \mathbf{q}_{n} + \Delta t V'\left(\mathbf{p}_{n+\nicefrac{1}{2}}\right),\label{eq:leapfrog_2}\\
    \mathbf{p}_{n+1} &= \mathbf{p}_{n+\nicefrac{1}{2}} - \nicefrac{1}{2}\Delta t T'\left(\mathbf{q}_{n+1}\right).\label{eq:leapfrog_3}
\end{align}
This scheme is central to SPar-GAN because it ensures that trajectories generated from the learned Hamiltonian are conservative over the full rollout horizon.
%
%
\subsection{Generative Adversarial Network}\label{subsec:gan}
Generative Adversarial Networks (GAN)~\citep{goodfellow2020generative} are a class of deep generative models that learn a joint distribution that can be sampled from and resembles the training dataset, without requiring an explicit model of the data distribution.
In the context of this paper, the training data consists of measured parachute trajectories, and the goal is to synthesize new trajectories that are statistically and physically consistent with the observed ones.

A GAN consists of two networks, a generator $G$ and a discriminator $D$.
The generator $G$ maps samples $z$ drawn from a simple distribution $p_{z}(z)$, typically a standard Gaussian, to synthetic data samples intended to resemble the training distribution $p_{\mathrm{data}}(x)$.
The discriminator $D$ takes a data sample as input and outputs the probability that the data sample came from the real dataset rather than from the generator.
The two networks are trained simultaneously with opposing objectives; the discriminator works to correctly classify real and synthetic samples, and the generator works to produce samples that the discriminator misclassifies as real.
This adversarial dynamic drives the generator toward producing samples that are progressively harder to distinguish from real data.
The objective takes the form of the two-player min max game
\begin{equation}
    \min_{G}\max_{D}\mathcal{L}\left(G, D\right) = \mathbb{E}_{z\sim p_{z}}\big[\log\left(1-D\left(G\left(z\right)\right)\right)\big] + \mathbb{E}_{x\sim p_{\mathrm{data}}}\big[\log\left(D\left(x\right)\right)\big].\label{eqn:gan_loss}
\end{equation}
This loss is maximized when the discriminator can perfectly distinguish real and generated samples, and minimized when the generator can produce generated data that is indistinguishable from the real data.
At this equilibrium, the discriminator output converges to $0.5$ for all real and generated data samples, and the generator's output distribution matches the training distribution.
After training, the discriminator is discarded and the generator alone is used to synthesize new samples by drawing $z$ from $p_{z}$ and passing it through $G$.

The GAN framework can be extended to condition generated samples on auxiliary information~\citep{mirza2014conditional}, such as parachute canopy design or freestream velocity in our setting.
This makes it possible to request samples corresponding to a specific configuration rather than drawing from the full distribution indiscriminately.
Concretely, both the generator and discriminator receive the conditioning variable $\xi$ as an additional input, and the objective becomes
\begin{equation}
    \min_{G}\max_{D}\mathcal{L}\left(G, D\right) = \mathbb{E}_{z\sim p_{z}}\big[\log\left(1-D\left(G\left(z|\xi\right)\right)\right)\big] + \mathbb{E}_{x\sim p_{\mathrm{data}}}\big[\log\left(D\left(x|\xi\right)\right)\big].
\end{equation}
At equilibrium, the generator learns to produce samples whose distribution matches $p_{\mathrm{data}}\left(x|\xi\right)$ for each configuration of $\xi$.
In this work this corresponds to learning the trajectory distribution associated with a specific canopy and flow condition.
%
\section{Generating Physically Plausible Parachute Trajectories}\label{sec:SParGAN}
\begin{figure}[t]
    \centering
    \includegraphics[width=0.9\linewidth]{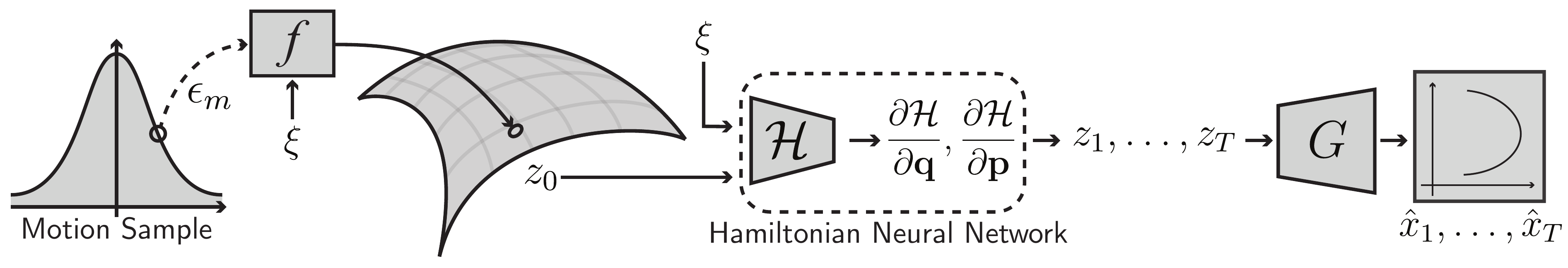}
    \caption{SPar-GAN for parachute trajectory generation. A random motion sample $\epsilon_{m}$ is passed through the configuration space map $f$ to produce an initial condition $z_{0}$ on the parachute phase-space. This initial condition is propagated forward in time by a Hamiltonian Neural Network conditioned on the system parameters $\xi$, yielding a phase-space trajectory $\mathbf{z}=\left[\mathbf{z}_{0},\mathbf{z}_{1},\ldots,\mathbf{z}_{T}\right]$. The generator $G$ then decodes this trajectory into the Euler-angle coordinates $\hat{\mathbf{x}}=\left[\hat{\mathbf{x}}_{0},\hat{\mathbf{x}}_{1},\ldots,\hat{\mathbf{x}}_{T}\right]$. The discriminator $D$ is omitted from this illustration for clarity.}
    \label{fig:ADS26_IntroCartoon}
    \vspace{-3mm}
\end{figure}
Given a dataset of parachute trajectories, our goal is to generate synthetic trajectories that are distinct from those in the training dataset but consistent in terms of the underlying physics.
SPar-GAN is a conditional generative model comprising three components trained end-to-end: a configuration space map, a phase-space motion model, and a generator (see Figure~\ref{fig:ADS26_IntroCartoon}).
The configuration space map (Section~\ref{subsec:configuration_space_map}) maps random Gaussian noise onto the parachute phase-space to provide an initial condition.
From there, the phase-space motion model (Section~\ref{subsec:phase_space_motion_model}) propagates the initial condition forward in time using a Hamiltonian Neural Network; this yields a sequence of states on the learned phase-space.
Finally, the generator (Section~\ref{subsec:generator}) decodes the trajectory into the Euler-angle coordinates; the same representation as the input data.
All three components are trained jointly under an adversarial objective augmented with a sparsity loss, which allows SPar-GAN to reproduce accurate parachute dynamics while recovering a physically meaningful phase-space.
%
%
\subsection{Configuration Space Map}\label{subsec:configuration_space_map}
The configuration space map $f:\mathbb{R}^{d_{\mathrm{motion}}}\times\mathbb{R}^{d_{\mathrm{label}}}\rightarrow\mathbb{R}^{2\times d_{\mathrm{lat}}}$ maps a Gaussian motion sample $\epsilon_{\mathrm{m}}\sim\mathcal{N}\left(0, \mathbf{I}_{d_{\mathrm{motion}}}\right)$ along with system configuration labels $\xi\in\mathbb{R}^{d_{\mathrm{label}}}$ (e.g., canopy type and wind tunnel velocity) to an initial condition for the phase-space motion model
\begin{equation}
    \mathbf{z}_{0} = f\left(\epsilon_{\mathrm{m}}, \xi\right) \in \mathbb{R}^{2\times d_{\mathrm{lat}}}.
\end{equation}
The output $\mathbf{z}_{0}=\left(\mathbf{q}_{0}, \mathbf{p}_{0}\right)$ is interpreted as a position-conjugate momentum pair on the learned phase-space.
Because the true dimension of the parachute phase-space is not known in advance, the dimension of the latent space $d_{\mathrm{lat}}$ is chosen so that it is at least as large as the expected number of degrees of freedom.
To encourage the model to recover the correct degrees of freedom, we incorporate a sparsity-inducing loss on the time derivatives of the momenta~\citep{allen2024hamiltonian,liu2025physically},
\begin{equation}
    \mathcal{L}_{\mathrm{cyclic}} = \lambda_{\mathrm{cyclic}}\sum\lvert \dot{p}_{i}\rvert.\label{eq:cyclic_coordinate_loss}
\end{equation}
Recall from Section~\ref{subsec:hamilton_mechanics} that cyclic coordinates have conserved conjugate momenta such that $\dot{p}_{i}=0$.
Without this penalty, the model would tend to spread its representation across all $d_{\mathrm{lat}}$ available dimensions.
Penalizing the magnitude of $\dot{p}_{i}$ instead encourages the representation to be compressed onto as few coordinates as the dynamics actually require, with the unused coordinates driven toward cyclic behavior.
The coefficient $\lambda_{\mathrm{cyclic}}$ controls the strength of this penalty, and allows SPar-GAN to recover the effective dimensionality of the parachute phase-space directly from data.
%
%
\subsection{Phase-Space Motion Model}\label{subsec:phase_space_motion_model}
Given the initial condition $\mathbf{z}_{0}$ from the configuration space map, the phase-space motion model rolls out the full trajectory $\mathbf{z}=\left[\mathbf{z}_{0},\mathbf{z}_{1},\ldots,\mathbf{z}_{T}\right]$ for time horizon $t=\left[0,\ldots,T\right]$.
Motion on the learned phase-space is governed by Hamilton's equations, where the Hamiltonian, $\mathcal{H}\left(\mathbf{q}, \mathbf{p}, \xi\right)$, is parameterized by an MLP.

A distinguishing feature of SPar-GAN is that the learned Hamiltonian is explicitly conditioned on the system labels $\xi$, the same system configuration provided to the configuration space map.
Rather than training a separate model for each configuration of canopy and flow condition, we expose $\xi$ as an input to the Hamiltonian network so that a single MLP represents a family of Hamiltonians indexed by $\xi$.
At inference time, specifying a value of $\xi$ selects a specific Hamiltonian, which in turn defines a specific parachute configuration.

Given the conditioned Hamiltonian, the required partial derivatives $\frac{\partial\mathcal{H}}{\partial\mathbf{q}}$ and $\frac{\partial\mathcal{H}}{\partial\mathbf{p}}$ are obtained through automatic differentiation of $\mathcal{H}\left(\mathbf{q}, \mathbf{p}, \xi\right)$.
System states are propagated forward in time using the symplectic leapfrog scheme described in Equations~\eqref{eq:leapfrog_1}-\eqref{eq:leapfrog_3}.
The resulting trajectory rollout is described by
\begin{equation}
    \mathbf{z}_{t+1} = \mathbf{z}_{t} + \int_{t}^{t+1}\dot{\mathbf{z}}_{\tau}\,d\tau = \mathbf{z}_{t} + \int_{t}^{t+1}\left(\frac{\partial\mathcal{H}\left(\mathbf{q}_{\tau}, \mathbf{p}_{\tau}, \xi\right)}{\partial\mathbf{p}_{\tau}}, -\frac{\partial\mathcal{H}\left(\mathbf{q}_{\tau}, \mathbf{p}_{\tau}, \xi\right)}{\partial\mathbf{q}_{\tau}}\right)\,d\tau.
\end{equation}
%
%
\subsection{Trajectory Generation}\label{subsec:generator}
As the phase-space trajectory $\mathbf{z}=\left[\mathbf{z}_{0},\mathbf{z}_{1},\ldots,\mathbf{z}_{T}\right]$ is not directly observable, SPar-GAN decodes these latent coordinates into Euler-angles to match the wind tunnel measurements of the real data.
This is done using the state generator $G:\mathbb{R}^{2\times d_{\mathrm{lat}}}\rightarrow\mathbb{R}^{d_{\mathrm{measure}}}$, where $d_{\mathrm{measure}}$ is the dimension of the Euler-angle measurements available from the wind tunnel photogrammetry.
The generator 
\begin{equation}
    \hat{\mathbf{x}}_{t} = G\left(\mathbf{z}_{t}\right)\in\mathbb{R}^{d_{\mathrm{measure}}},
\end{equation}
is an MLP applied pointwise to each state along the phase-space trajectory, yielding the synthetic trajectory $\hat{\mathbf{x}}=\left[\hat{\mathbf{x}}_{0},\hat{\mathbf{x}}_{1},\ldots,\hat{\mathbf{x}}_{T}\right]$.

To ensure synthetic trajectories are statistically indistinguishable from real ones, SPar-GAN uses a discriminator $D$ that takes trajectories and system labels $\xi$ as input, and outputs the probability that the trajectory is real.
The discriminator is parameterized by a recurrent neural network~\citep{jordan1997serial} to capture temporal structure across the trajectory.
The generator and discriminator are trained under adversarial loss
\begin{equation}
    \mathcal{L}_{\mathrm{traj}} = \mathbb{E}_{\mathbf{x}\sim p_{X}}\left[\log D\left(\mathbf{x}\right)\right] + \mathbb{E}_{\hat{\mathbf{x}}\sim p_{G}}\left[\log \left(1-D\big(\hat{\mathbf{x}}\big)\right)\right].
\end{equation}
The full training objective combines this adversarial loss with the cyclic coordinate loss from Section~\ref{subsec:configuration_space_map},
\begin{equation}
    \mathcal{L} = \mathcal{L}_{\mathrm{traj}} + \mathcal{L}_{\mathrm{cyclic}},
\end{equation}
The two losses play complementary roles. 
The adversarial loss $\mathcal{L}_{\mathrm{traj}}$ enforces dynamical accuracy by requiring generated trajectories to be indistinguishable from measured ones.
The cyclic coordinate loss $\mathcal{L}_{\mathrm{cyclic}}$ limits the effective dimension of the latent phase-space, allowing SPar-GAN to recover the underlying dynamical structure rather than fitting the data with an unnecessarily large representation.
%
\section{Wind Tunnel Dataset}\label{sec:dataset}
%
\begin{figure}[b]
    \centering
    \includegraphics[width=\linewidth]{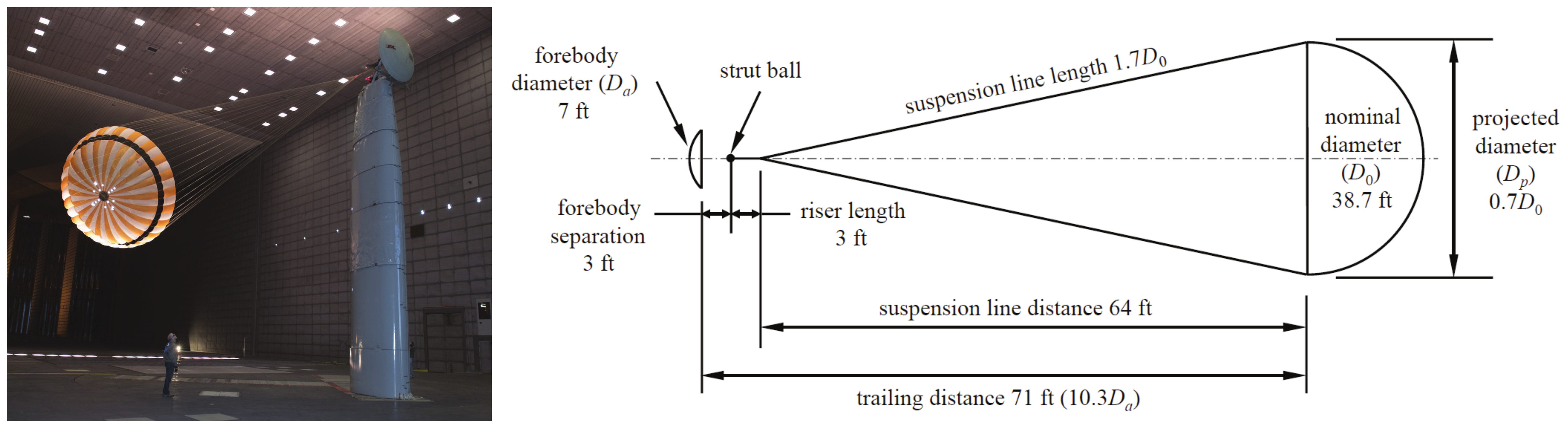}
    \vspace{-4mm}
    \caption{Schematic of the LDSD subscale test conducted at NFAC. (Left) Photo of a subscale ringsail parachute in the NFAC wind-tunnel, showing the $\mathbf{40}$ ft strut with fairing and forebody in the $\mathbf{80\times120}$ test section. (Right) Major dimensions of the model support system and test articles. All indicated dimensions are approximate.}
    \label{fig:SPSGAN_ADS26_NFACSchematic}
    \vspace{-3mm}
\end{figure}
\begin{figure}[t]
    \vspace{-4mm}
    \centering
    \includegraphics[width=0.9\linewidth]{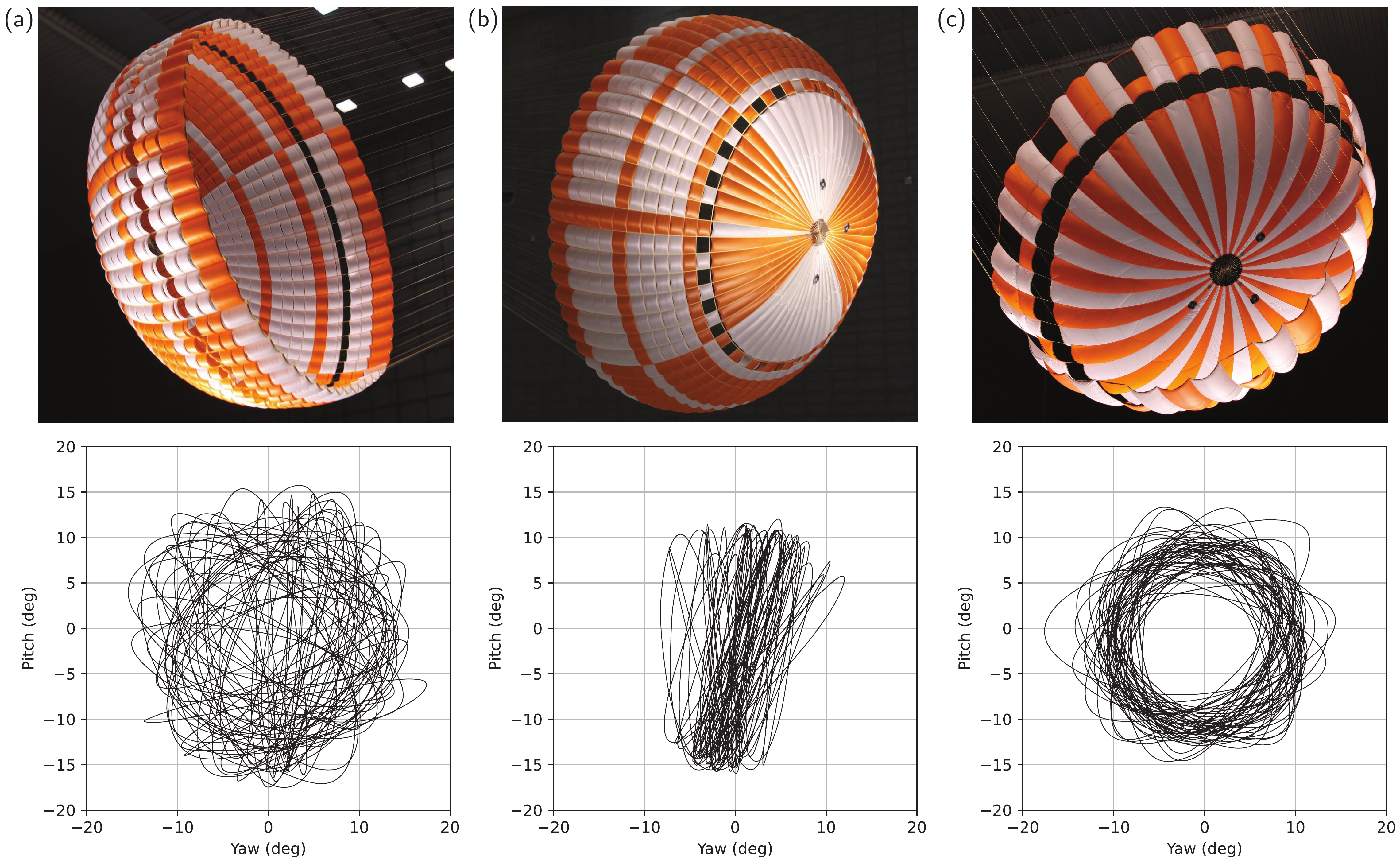}
    \vspace{-3mm}
    \caption{Image and example trajectories of studied canopies. (a) Ringsail canopy; (b) Disksail canopy; (c) Modified DGB canopy. (Top) Image of the canopy; (Bottom) Trajectory of the canopy apex.}
    \label{fig:SPSGAN_ADS26_NFACSchematicComposite}
    \vspace{-3mm}
\end{figure}
\begin{figure}[b]
    \vspace{-4mm}
    \centering
    \includegraphics[width=0.35\linewidth]{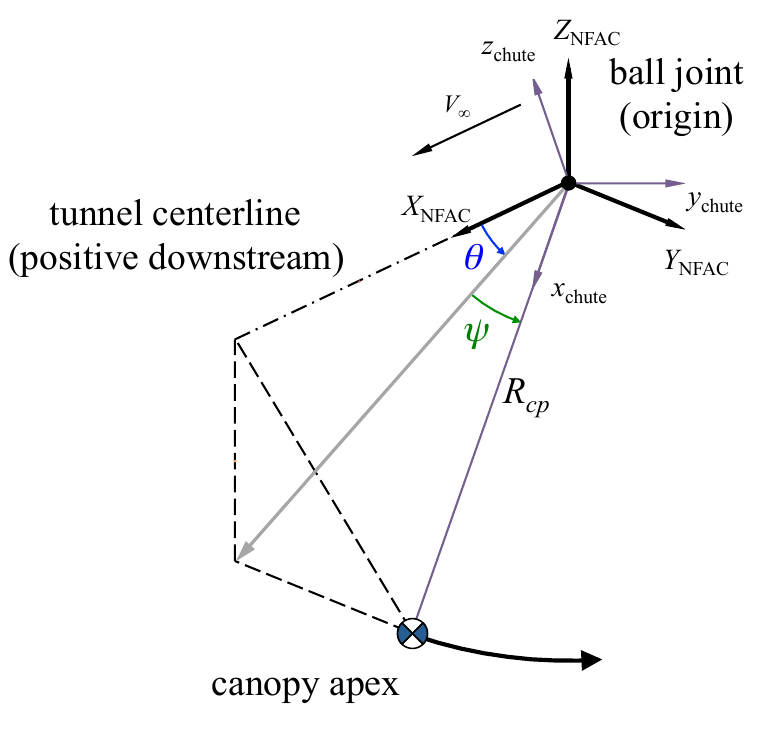}
    \vspace{-3mm}
    \caption{NFAC wind tunnel ($X_{NFAC}$, $Y_{NFAC}$, $Z_{NFAC}$) and parachute ($x$, $y$, $z$) coordinate systems. Both coordinate frames are centered at the test fixture ball joint. The $X_{NFAC}$ axis is aligned with the tunnel centerline, along the direction of the freestream velocity $V_\infty$; the $Z_{NFAC}$ axis points towards the ceiling. The $x$ axis is aligned with the parachute centerline and points towards the canopy apex. The Euler pitch $\theta$ and yaw $\psi$ angles are defined as shown.  }
    \label{fig:coordsys}
    \vspace{-3mm}
\end{figure}
We study subscale parachutes~\citep{tanner2013aerodynamic} tested at the National Full Scale Aerodynamics Complex (NFAC) as part of the Low-Density Supersonic Decelerator (LDSD) project~\citep{clark2013development}.
The test setup and dimensions of the test articles are shown in Figure~\ref{fig:SPSGAN_ADS26_NFACSchematic}. The parachutes were fixed to a strut at the center of the 80-by-120-foot test section by a load arm and ball joint which allowed the test articles to rotate freely.
Five different canopy configurations of ringsail and ringsail-derived parachute designs were tested at sea-level air density and freestream speeds ranging from approximately 15 knots (7.7~m/s) to 30 knots (15.4~ m/s).
The ringsail models had a nominal diameter ($D_0$) of 11.8~m and were constructed using PIA-C-44378D Type~I nylon, which has near-zero permeability. 
The geometric porosity (the ratio of open area to nominal area) of the canopies ranged from approximately 10\% to 22\%.
Additional Disk-Gap-Band (DGB) canopies of the same nominal diameter were tested to obtain reference performance criteria. Throughout the test, the standard ringsail, disksail, and DGB canopies were subjected to small modifications (eg: removal of fabric in specific regions) to observe the effect on system dynamics.
Figure~\ref{fig:SPSGAN_ADS26_NFACSchematicComposite} shows three test articles from the test campaign: a ringsail, a disksail, and a modified DGB.

The tangential force generated by the parachutes were measured by a load cell at the attachment of the riser to the test fixture and recorded at 1000~Hz. Retro-reflective targets were attached to the canopies, and motion of the canopies was reconstructed using a photogrammetry system consisting of three synchronized cameras recording at $60$ fps. The position and orientation of the canopy in a coordinate system fixed to the wind tunnel (Figure~\ref{fig:coordsys}) were reconstructed as described in Reference~\cite{schairer2018measurements}. Example trajectories of the canopy apex are shown in Figure~\ref{fig:SPSGAN_ADS26_NFACSchematicComposite}. 

We aim to generate synthetic trajectories that are physically consistent with the data collected at NFAC. To this end, we consider three canopy configurations:
\begin{itemize}
    \item Ringsail canopy, shown in Figure~\ref{fig:SPSGAN_ADS26_NFACSchematicComposite}(a), flown at $30$ knots freestream velocity. Its dynamics are characterized by oscillations of similar magnitude in both pitch and yaw.
    \item Disksail canopy flown at $25$ knots freestream velocity. This canopy, shown in Figure~\ref{fig:SPSGAN_ADS26_NFACSchematicComposite}(b), was modified from a ringsail by replacing with rings 1 through 11 with a flat circular disk. Its dynamics are dominated by motion in the pitch axis.
    \item Modified DGB canopy flown at $30$ knots freestream velocity. As shown in Figure~\ref{fig:SPSGAN_ADS26_NFACSchematicComposite}(c), this canopy was modified from the standard DGB by removing fabric from every third gore on the cylindrical band. Its dynamics trace an orbital motion on the pitch-yaw phase plane.
\end{itemize}
For all tasks reported in Section~\ref{sec:results}, SPar-GAN is trained for 20,000 iterations on real trajectory segments obtained from NFAC wind tunnel testing, obtained by downsampling the 60 Hz photogrammetry measurements to 30 Hz and extracting 30-frame/1-second windows.
For both real training and generated trajectories, random 16-frame partitions are passed to the discriminator when training the model.
At inference time, the generator produces synthetic trajectories of the same 30-frame length.
The generated trajectories are evaluated on their
\begin{itemize}
    \item Consistency. The distribution of generated trajectories is compared against the distribution of real trajectories, to assess whether the generated data evolves within the same envelope as the measurements.
    \item Phase-space dimensionality. The latent phase-space has dimension $d_{\mathrm{lat}}$, set larger than the expected number of degrees of freedom and reduced during training by the cyclic-coordinate loss in Equation~\eqref{eq:cyclic_coordinate_loss}. To estimate the resulting effective dimensionality, we apply Principal Component Analysis (PCA) to the latent states $\left[\mathbf{z}_{0},\mathbf{z}_{1},\ldots,\mathbf{z}_{T}\right]$ collected from the generated trajectories. The eigenvalues of the covariance matrix give the variance captured along each principal direction; their relative magnitudes indicate which directions are dynamically active and which have been driven toward cyclic coordinates. We then apply Cattell's criterion~\citep{cattell1966scree} to identify the elbow in this spectrum and read off the effective number of degrees of freedom.
\end{itemize}
%
\section{Results}\label{sec:results}
%
\subsection{Ringsail Canopy}\label{subsec:ringsail_parachute}
\begin{figure}[b]
    \centering
    \includegraphics[width=\linewidth]{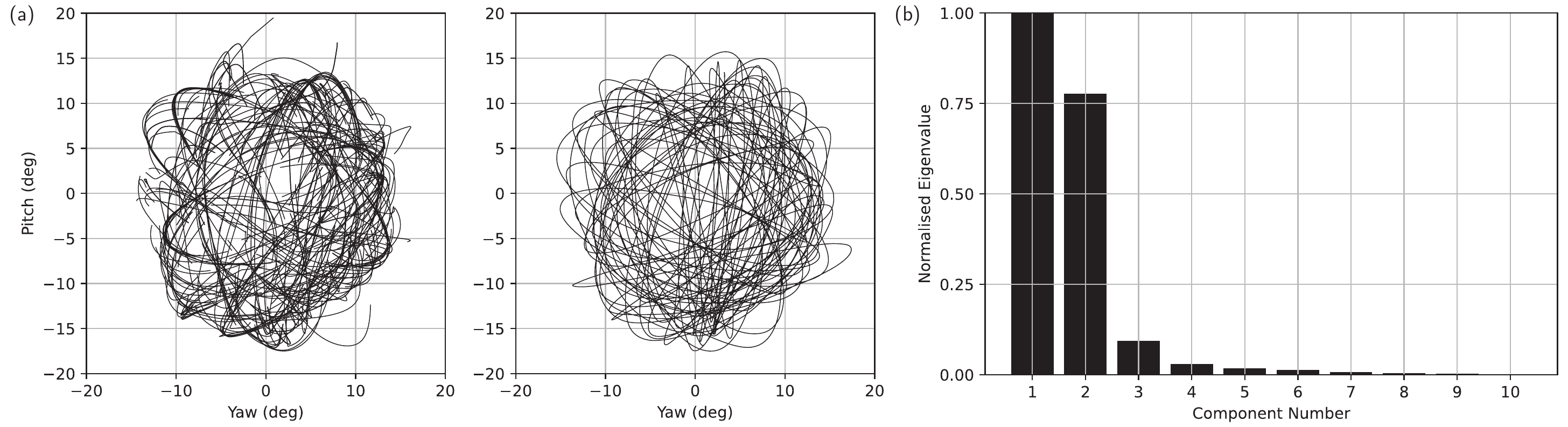}
    \vspace{-6mm}
    \caption{SPar-GAN generated trajectories for the ringsail canopy. (a) Comparison between the distribution of generated trajectories (left) and the distribution of real data (right). (b) PCA eigenvalue analysis of the learned phase-space, normalized by the largest component.}
    \label{fig:SParGAN_ADS26_NFACComposite_RS1}
    \vspace{-3mm}
\end{figure}
We train and test SPar-GAN on the ringsail canopy shown in Figure~\ref{fig:SPSGAN_ADS26_NFACSchematicComposite}(a).
After training, we draw 1000 synthetic trajectories from the generator for evaluation.
The distribution of the generated trajectories is shown alongside the real data distribution in Figure~\ref{fig:SParGAN_ADS26_NFACComposite_RS1}(a).
The two distributions occupy similar envelopes on the pitch-yaw plane, with oscillations of comparable amplitude in both the pitch and yaw axis, indicating that SPar-GAN reproduces the characteristic coupled pitch-yaw motion of the ringsail canopy.
PCA eigenvalue analysis of the learned phase-space, shown in Figure~\ref{fig:SParGAN_ADS26_NFACComposite_RS1}(b), identifies a two-degree-of-freedom structure based on Cattell's criterion~\citep{cattell1966scree}.
This is consistent with the axisymmetry of the parachute, for which the canopy position can be approximated using only its pitch and yaw.
%
%
\subsection{Disksail Canopy}\label{subsec:disksail_parachute}
\begin{figure}[t]
    \centering
    \includegraphics[width=\linewidth]{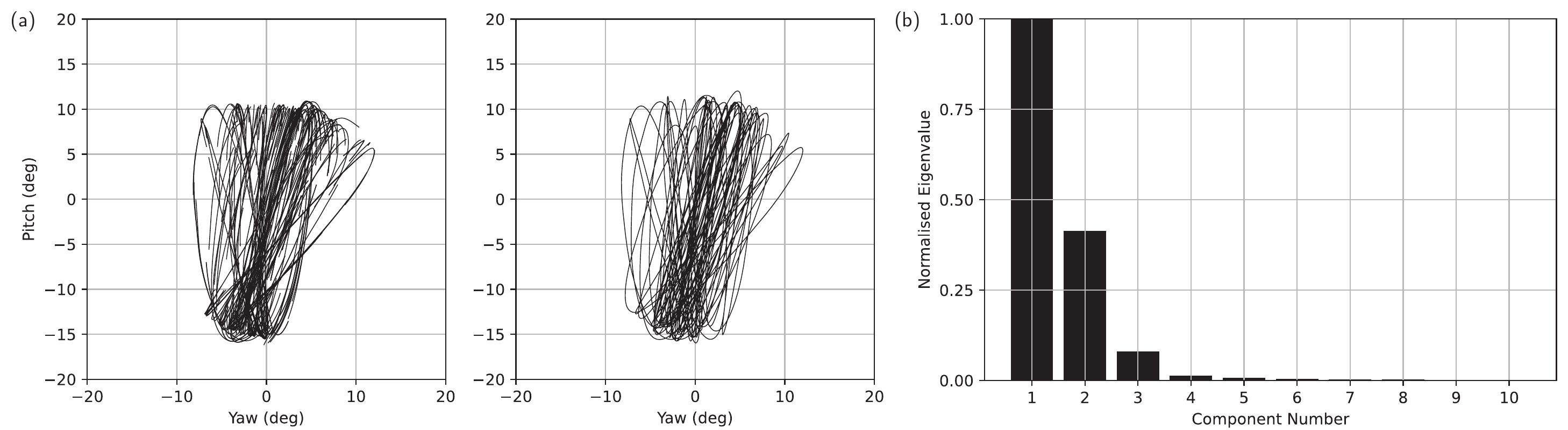}
    \vspace{-6mm}
    \caption{SPar-GAN generated trajectory for the disksail canopy. (a) Comparison between the distribution of generated trajectories (left) and the distribution of real data (right). (b) PCA eigenvalue analysis of the learned phase-space, normalized by the largest component.}
    \label{fig:SParGAN_ADS26_NFACComposite_DS1}
    \vspace{-3mm}
\end{figure}

We train and test SPar-GAN on the disksail canopy shown in Figure~\ref{fig:SPSGAN_ADS26_NFACSchematicComposite}(b).
After training, we draw 1000 synthetic trajectories from the generator for evaluation.
The distribution of the generated trajectories is shown alongside the real data distribution in Figure~\ref{fig:SParGAN_ADS26_NFACComposite_DS1}(a).
The two distributions occupy similar envelopes on the pitch-yaw plane, with oscillations that extend far more in pitch than in yaw, indicating that SPar-GAN reproduces the characteristic pitch-dominated motion of the disksail canopy.
PCA eigenvalue analysis of the learned phase-space, shown in Figure~\ref{fig:SParGAN_ADS26_NFACComposite_DS1}(b), identifies a two-degree-of-freedom structure based on Cattell's criterion.
The magnitude of the second principal component is much smaller than for the ringsail canopy in Section~\ref{subsec:ringsail_parachute}, reflecting the fact that the disksail canopy exhibits much less motion in the yaw axis than in pitch.
%
%
\subsection{Modified Disk-Gap-Band Canopy}
\begin{figure}[b]
    \centering
    \includegraphics[width=\linewidth]{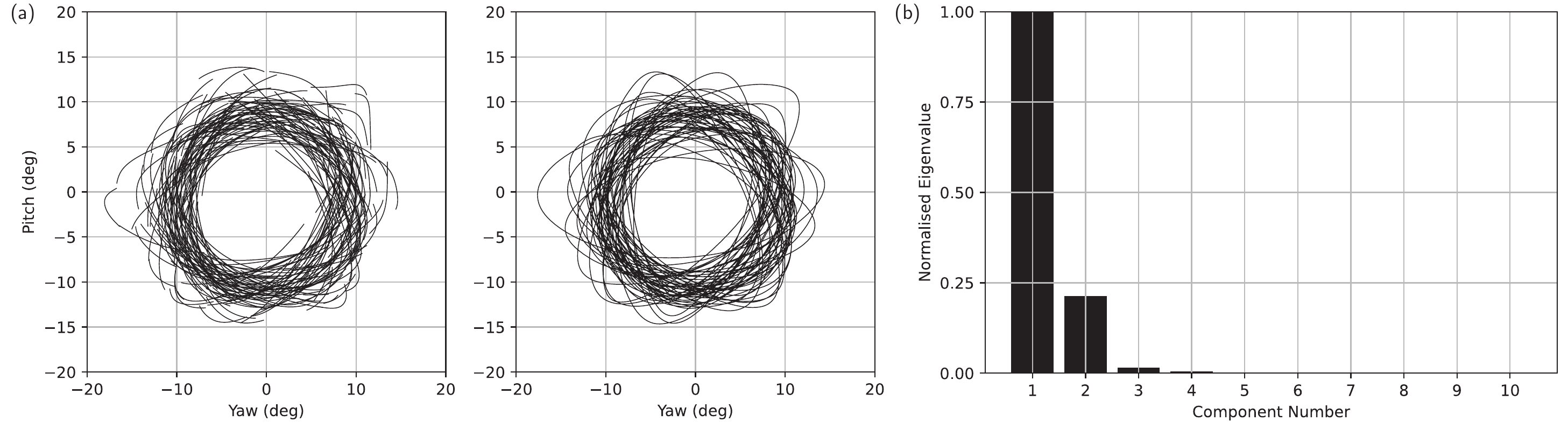}
    \vspace{-6mm}
    \caption{SPar-GAN generated trajectory for the modified DGB canopy. (a) Comparison between the distribution of generated trajectories (left) and the distribution of real data (right). (b) PCA eigenvalue analysis of the learned phase-space, normalized by the largest component.}
    \label{fig:SParGAN_ADS26_NFACComposite_DGB3}
    \vspace{-3mm}
\end{figure}
\begin{figure}[t]
    \centering
    \includegraphics[width=0.6\linewidth]{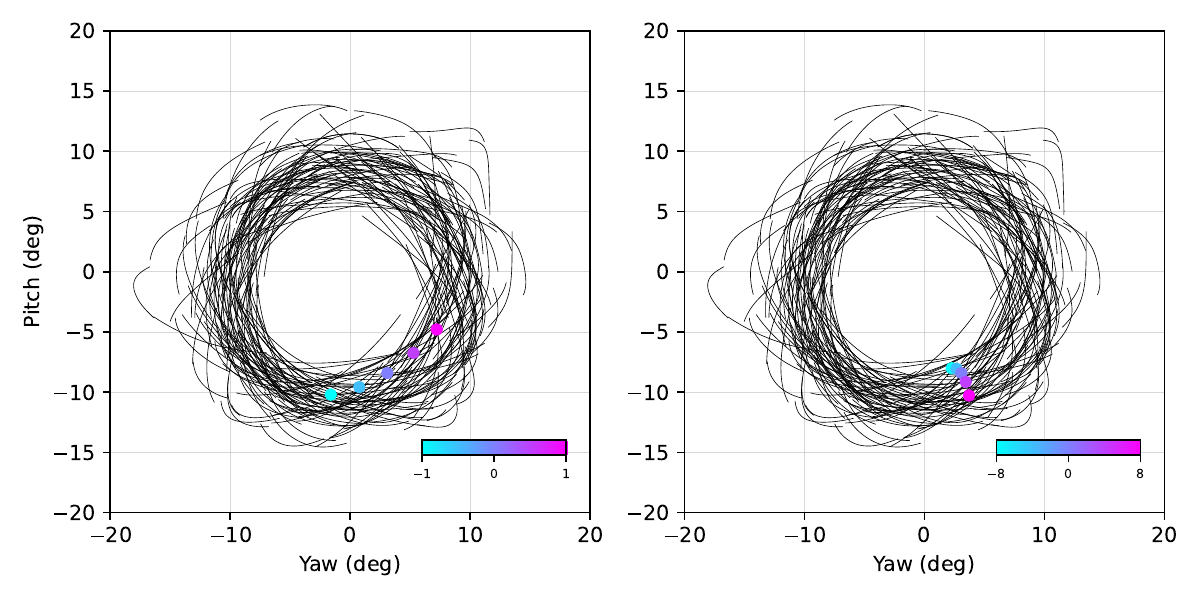}
    \vspace{-3mm}
    \caption{SPar-GAN identified canonical coordinates for modified DGB. Five points equally spaced between $-1\sigma/8\sigma$ and $+1\sigma/8\sigma$ along each principal component is decoded into Euler-angle representations. (Left) Principal component 1 accounts for 80.6\% of the variance explained. The decoded trajectories sweep along the approximate angular direction of the orbit - the trajectory shape is preserved while its phase/starting position rotates around the centroid. (Right) Principal component 2 accounts for 17.2\% of the variance explained. The decoded trajectories move radially - changing the distance from the centroid while preserving the phase.}
    \label{fig:SParGAN_ADS26_DGBPCAProjection}
    \vspace{-3mm}
\end{figure}
We train and test SPar-GAN on the modified DGB canopy shown in Figure~\ref{fig:SPSGAN_ADS26_NFACSchematicComposite}(c).
After training, we draw 1000 synthetic trajectories from the generator for evaluation.
The distribution of the generated trajectories is shown alongside the real data distribution in Figure~\ref{fig:SParGAN_ADS26_NFACComposite_DGB3}(a).
The two distributions occupy similar envelopes on the pitch-yaw plane, with trajectories that trace ring-shaped orbits around the origin, indicating that SPar-GAN reproduces the orbital motion of the DGB canopy.
PCA eigenvalue analysis of the learned phase-space, shown in Figure~\ref{fig:SParGAN_ADS26_NFACComposite_DGB3}(b), identifies a two-degree-of-freedom structure based on Cattell's criterion.
The magnitude of the second principal component is much smaller than for both the ringsail canopy in Section~\ref{subsec:ringsail_parachute} and the disksail canopy in Section~\ref{subsec:disksail_parachute}.
Since the motion is orbital, a plausible explanation is that the system can be described using log-polar coordinates, with the first principal component capturing the angle and the second capturing the radius. 
The deviation in radius is significantly smaller than the deviation in angle, so the second principal component contributes less to the overall variance.
We empirically verify this intuition in Figure~\ref{fig:SParGAN_ADS26_DGBPCAProjection}.
From a reference point on the learned phase-space, we take five samples equally spaced along each of the first two principal directions, between $\pm1\sigma$ along the first component and between $\pm8\sigma$ along the second, and decode each into Euler-angle coordinates.
Samples along the first principal component produce orbits whose phase rotates around the centroid while the overall ring shape is preserved, corresponding to angular motion along the ring.
Samples along the second principal component produce orbits that shift radially inward or outward from the centroid while the phase is preserved, corresponding to radial motion.
These decoded motions match the expected behavior under a log-polar representation.
%
%
\subsection{Conditional Interpolation and Extrapolation of Ringsail Canopy}
\begin{figure}[b]
    \vspace{-3mm}
    \centering
    \begin{subfigure}[t]{0.49\textwidth}
        \centering
        \includegraphics[width=\textwidth]{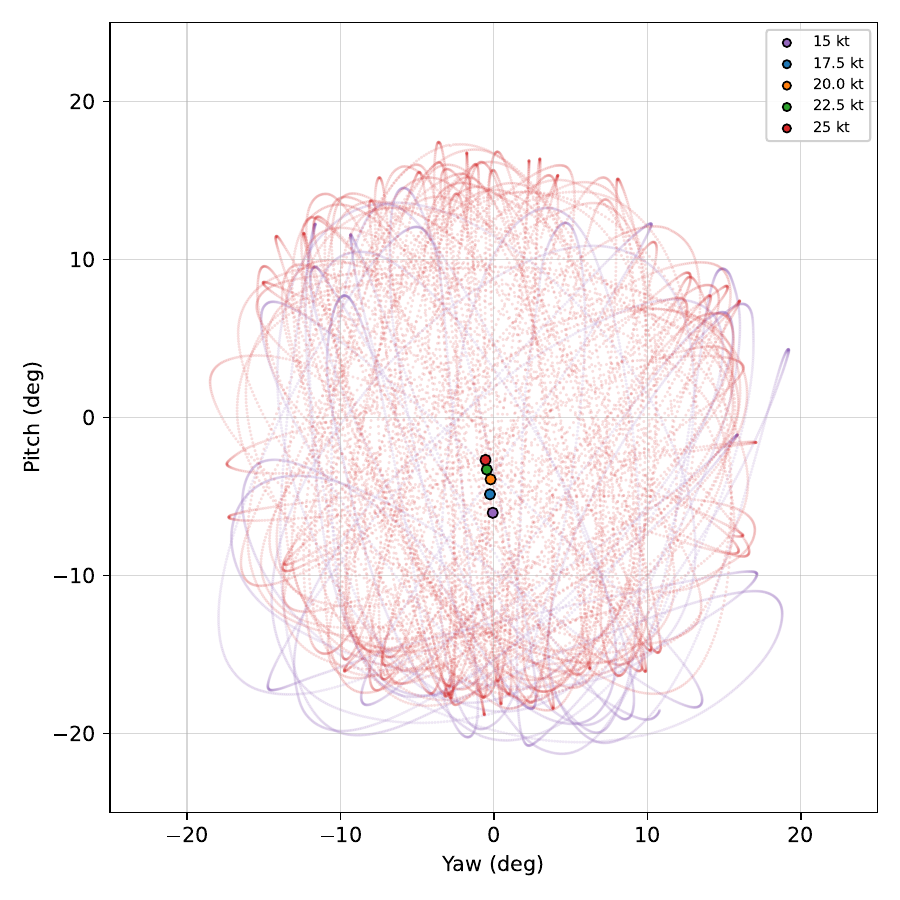}
        \captionsetup{width=0.9\textwidth}
        \caption{Interpolation under wind tunnel freestream velocity of 17.5, 20.0, and 22.5 knots.}
        \label{fig:SparGAN_ADS26_ConditionalInterpolateCoM}
    \end{subfigure}
    \begin{subfigure}[t]{0.49\textwidth}
        \centering
        \includegraphics[width=\textwidth]{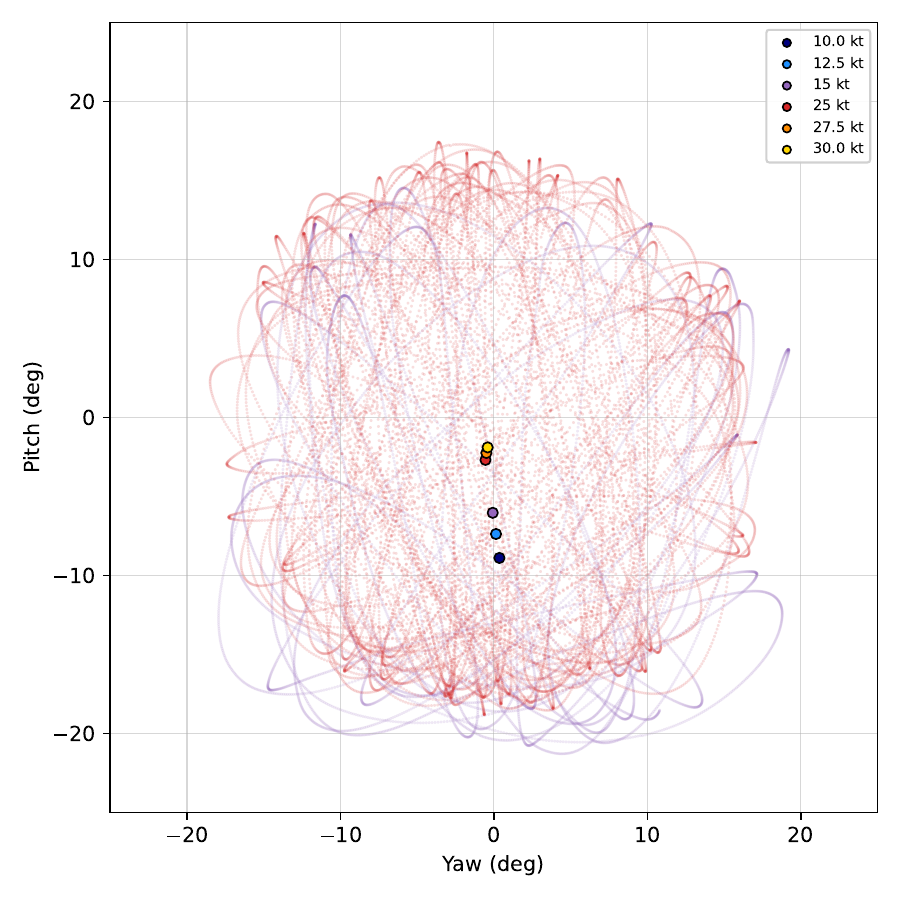}
        \captionsetup{width=0.9\textwidth}
        \caption{Extrapolation under wind tunnel freestream velocity of 10.0, 12.5, 27.5, and 30.0 knots.}
        \label{fig:SparGAN_ADS26_ConditionalExtrapolateCoM}
    \end{subfigure}
    \caption{SPar-GAN generated trajectory for ringsail canopy under unseen wind tunnel conditions. (a) Shows the center of mass under freestream velocities interpolated between 15 and 25 knots; (b) Shows the center of mass under freestream velocities extrapolated beyond 15 and 25 knots.}
    \label{fig:spargan_conditional_generation}
\end{figure}
\begin{figure}[t]
    \centering
    \includegraphics[width=\linewidth]{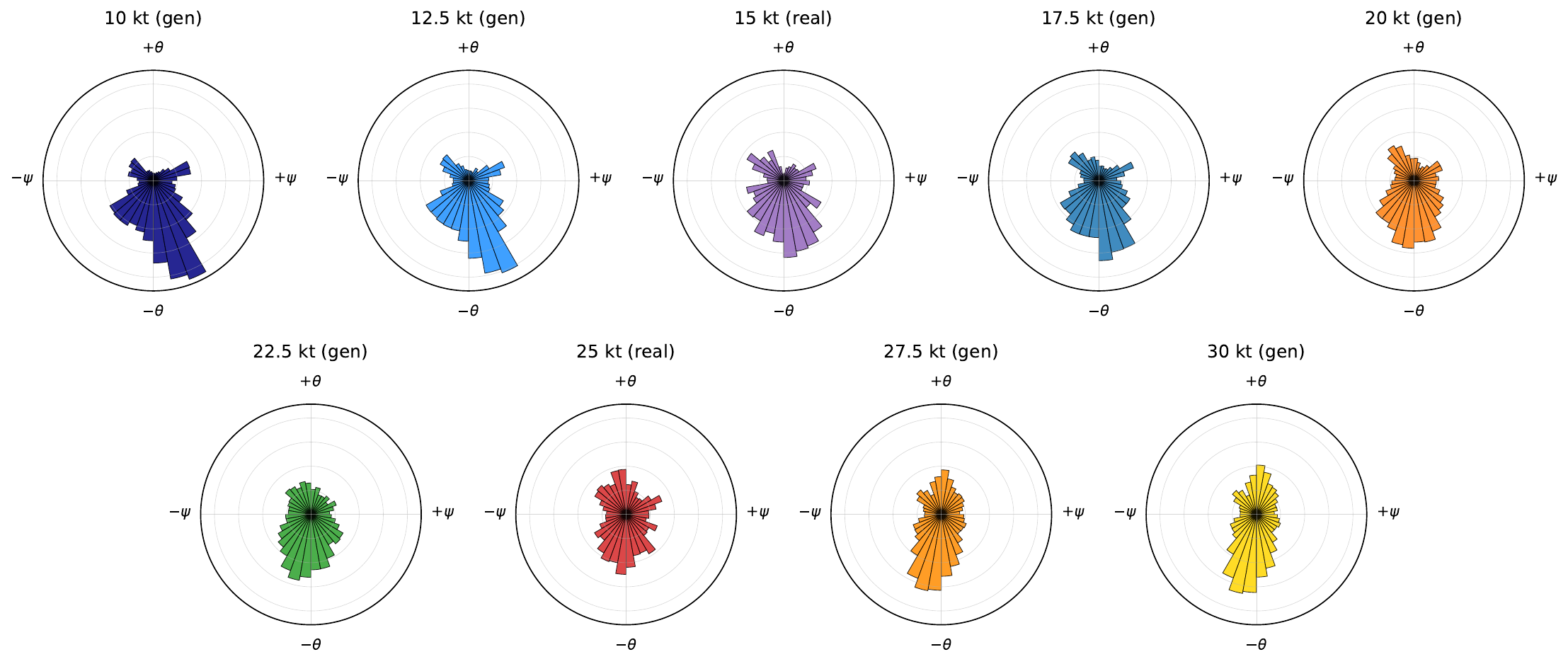}
    \vspace{-3mm}
    \caption{SPar-GAN generated distribution for ringsail canopy under unseen wind tunnel conditions. Each rose plot shows the pitch-yaw distribution of 1000 generated trajectories at the indicated velocity. Real training data is available only at 15 and 25 knots (labeled "real"); all other conditions are generated by conditioning on the corresponding freestream velocity.}
    \label{fig:SparGAN_ADS26_ConditionalInterExtraRose}
    \vspace{-3mm}
\end{figure}
A key benefit of generative modeling is the ability to produce novel trajectories that remain physically consistent with the training data.
SPar-GAN supports this by exposing the physical parameters $\xi$ as conditioning inputs, which allows the model to generate trajectories at operating conditions not represented in the training set simply by changing the desired value of $\xi$ at inference time.
To demonstrate this capability, we train SPar-GAN on ringsail canopy trajectories collected at 15 and 25 knots freestream velocity and evaluate it at both interpolated (17.5, 20.0, and 22.5 knots) and extrapolated velocities (10.0, 12.5, 27.5, and 30.0 knots).
The centers of mass (CoM) and trajectory distribution of the generated trajectories are shown in Figure~\ref{fig:spargan_conditional_generation} and Figure~\ref{fig:SparGAN_ADS26_ConditionalInterExtraRose} respectively.
Under interpolated conditions, the CoM transitions smoothly between the two training conditions, with the mean pitch decreasing monotonically as freestream velocity increases.
Under extrapolated conditions, the CoM continues this trend, pitching further upward at higher velocities and further downward at lower velocities.
The yaw-axis distribution also broadens further as velocity decreases below the training range.
%
%
\subsection{Conditional Transfer of Ringsail and Disksail Canopy}
\begin{figure}[t]
    \centering
    \includegraphics[width=\linewidth]{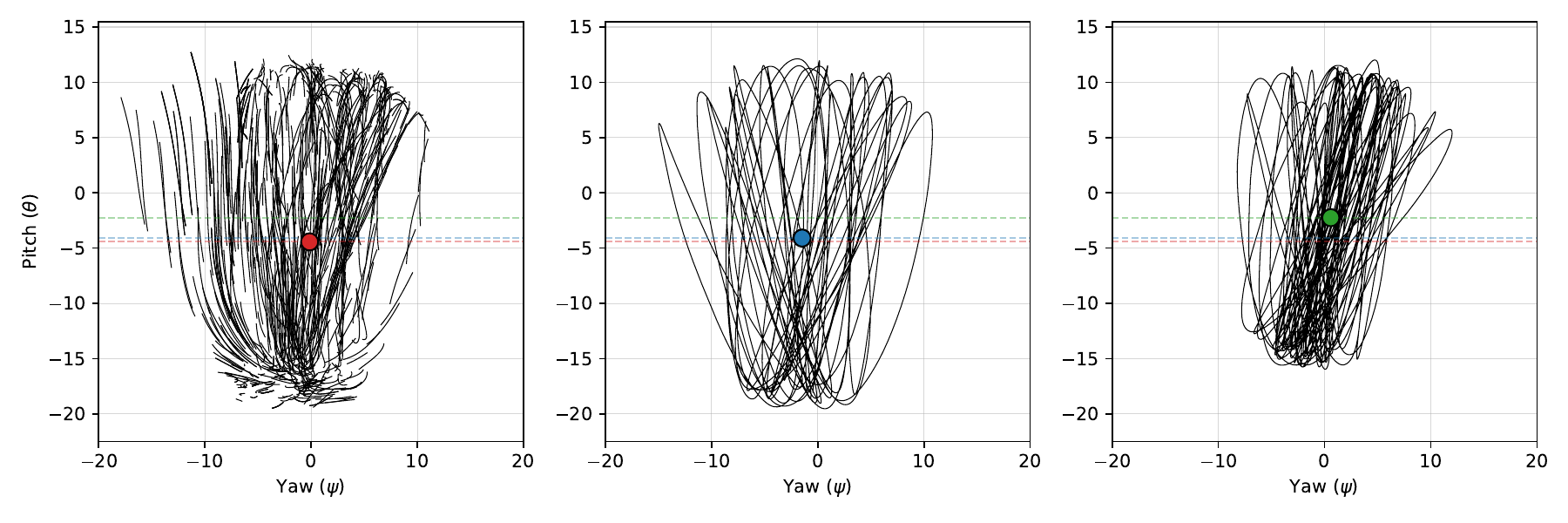}
    \vspace{-6mm}
    \caption{SPar-GAN dynamics transfer from ringsail to disksail at an unseen freestream velocity. SPar-GAN is trained on ringsail trajectories at 15 and 25 knots and disksail trajectories at 25 knots, then tasked with generating disksail trajectories at 15 knots. (Left) SPar-GAN generated trajectories at the unseen 15-knot condition; (Center) Real disksail trajectories at 15 knots, held out from training; (Right) Disksail training data at 25 knots. The colored dots and dashed lines indicate the CoM and pitch CoM respectively.}
    \label{fig:SparGAN_ADS26_ConditionalTransferCoM}
    \vspace{-3mm}
\end{figure}
A further application of the conditional framework is rapid prototyping of new canopy designs at untested operating conditions. 
We train SPar-GAN on ringsail canopies flown at 15 and 25 knots freestream velocity and disksail canopies flown at 25 knots.
During inference, SPar-GAN generates disksail trajectories at 15 knots, a combination of canopy and velocity that does not appear in the training set.
This setup tests whether SPar-GAN can compose the velocity dependence learned from the ringsail data with the canopy-specific dynamics learned from the single disksail condition.

The generated trajectories, shown in Figure~\ref{fig:SparGAN_ADS26_ConditionalTransferCoM}, exhibit a center of mass aligned with the unseen 15-knot reference rather than with the seen 25-knot reference.
Furthermore, the generated trajectory distribution is statistically closer to the unseen 15-knot ground truth than to the 25-knot training data. 
We quantify this using the Kullback-Leibler (KL) divergence~\citep{shlens2014notes}, a standard measure of the difference between two probability distributions, with smaller values indicating closer agreement. 
The generated distribution achieves a KL divergence of 1.23 against the 15-knot reference and 5.24 against the 25-knot training data, confirming that SPar-GAN matches the unseen condition more closely than the seen one.
This indicates that SPar-GAN has disentangled the effect of freestream velocity from the effect of canopy geometry, allowing it to extrapolate dynamics for new canopies at conditions where only a small amount of paired data is available.
%
\section{Conclusion}\label{sec:conclusion}
%
In this paper, we introduce SPar-GAN, a physics-constrained generative model for parachute trajectory dynamics, and demonstrate its capabilities on subscale canopies tested in the NFAC wind tunnel.
SPar-GAN generates trajectories that are highly consistent with real data across three qualitatively different canopy designs, recovering the uniform pitch-yaw motion of the ringsail, the pitch-dominated motion of the disksail, and the orbital motion of the modified DGB canopy.
For each canopy, the learned phase-space reduces to a two-degree-of-freedom structure consistent with canopy axisymmetry, and for the modified DGB canopy the principal directions correspond directly to the angular and radial components of the orbital motion.

Beyond reproducing observed dynamics, SPar-GAN generalizes to operating conditions not represented in the training set.
Conditioning on freestream velocity enables smooth interpolation between tested velocities and physically plausible extrapolation beyond them.
Conditioning jointly on velocity and canopy geometry allows SPar-GAN to transfer dynamics across canopy designs, generating trajectories for canopy-velocity combinations never observed in training.
These results suggest that physics-constrained generative models can serve as a practical tool for characterizing parachute dynamics across operating conditions and may help reduce the volume of physical testing required to assess performance.

Several directions remain open for future work. Incorporating load-cell force measurements as an additional conditioning input would allow SPar-GAN to generate trajectories that are consistent with a target aerodynamic loading profile. Extending the model to incorporate derivative measurements, such as angular rates and accelerations from inertial sensors, would provide a more complete description of the parachute state and enable SPar-GAN to capture higher-order dynamical features that are difficult to infer from position measurements alone. Together, these extensions would broaden the scope of SPar-GAN from trajectory generation toward a richer characterization of parachute behavior suitable for design and analysis across the full EDL envelope.
\section{Acknowledgments}
This research was carried out at the Jet Propulsion Laboratory, California Institute of Technology, under a contract with the National Aeronautics and Space Administration and funded through the internal Strategic University Research Partnerships program. The authors would like to gratefully acknowledge the contribution and discussion from Katherine J.Y. Siegel and Kirin Peterson. We thank Dr. Ian G. Clark and Dr. Christopher Tanner of JPL for access to the wind tunnel data set. 
\bibliography{ref}
\end{document}